\documentclass[journal]{IEEEtran}
\IEEEoverridecommandlockouts
\ifCLASSINFOpdf
\else
\fi
\usepackage{cite}
\usepackage{amsthm}
\usepackage{multirow}
\usepackage{balance}
\usepackage{latexsym}
\usepackage{graphicx}
\usepackage{soul,color}
\usepackage{amsmath}
\usepackage{algorithm}
\usepackage[noend]{algpseudocode}
\usepackage[utf8]{inputenc}
\usepackage[export]{adjustbox}
\usepackage[T1]{fontenc}
 \usepackage{array,multirow,graphicx}
 \usepackage{float}
 \usepackage{hyperref}
 \usepackage{subcaption}
 \usepackage{mathtools}
 \usepackage[normalem]{ulem}
 \usepackage{amssymb}

 \usepackage[table]{xcolor}

 \makeatletter
\def\BState{\State\hskip-\ALG@thistlm}
\makeatother

 \newcommand\blfootnote[1]{%
  \begingroup
  \renewcommand\thefootnote{}\footnote{#1}%
  \addtocounter{footnote}{-1}%
  \endgroup}

\begin{document}
\title{An Adaptive Multi-Agent Physical Layer Security Framework for Cognitive Cyber-Physical Systems}

\author{\IEEEauthorblockN{Mehmet Özgün Demir, Ozan Alp Topal, Ali Emre Pusane, Guido Dartmann, \\  Gerd Ascheid, and  Güneş Karabulut Kurt}
%\\
%\IEEEauthorblockA{{\IEEEauthorrefmark{2}Boğaziçi University, Istanbul, Turkey, \{ozgun.demir1, ali.pusane\}@boun.edu.tr}}\\
%\IEEEauthorblockA{{\IEEEauthorrefmark{3}RWTH Aachen University, Aachen, Germany, \{liang,ascheid\}@ice.rwth-aachen.de}}\\
%\IEEEauthorblockA{{\IEEEauthorrefmark{4}University of Applied Sciences Trier, Trier, Germany, 
%g.dartmann@umwelt-campus.de}}
 \thanks{ M. Ö. Demir and A. E. Pusane are with the Boğaziçi University, Istanbul, Turkey, \{ozgun.demir1, ali.pusane\}@boun.edu.tr.} 
 \thanks{O. A. Topal and G. Karabulut Kurt are with the Istanbul Technical University, Istanbul, Turkey, 
\{ozan.topal,gkurt\}@itu.edu.tr. }
\thanks{\noindent G. Dartmann is with the University of Applied Sciences Trier, Trier, Germany,
g.dartmann@umwelt-campus.de.} 
\thanks{\noindent G. Ascheid is with the RWTH Aachen University, Aachen, Germany, ascheid@ice.rwth-aachen.de.} }

% \markboth{IEEE Wireless Communications,~Vol.XX, No.XX, XXXX}%
% {Shell \MakeLowercase{\textit{et al.}}: IEEE Wireless Communications}

\maketitle

\begin{abstract}
Being capable of sensing and behavioral adaptation in line with their changing environments, cognitive cyber-physical systems (CCPSs) are the new form of applications in future wireless networks. With the advancement of the machine learning algorithms, the transmission scheme providing the best performance can be utilized to sustain a reliable network of CCPS agents equipped with self-decision mechanisms, where the interactions between each agent are modeled in terms of service quality, security, and cost dimensions. In this work, first, we provide network utility as a reliability metric, which is a weighted sum of the individual utility values of the CCPS agents. The individual utilities are calculated by mixing the quality of service (QoS), security, and cost dimensions with the proportions determined by the individualized user requirements. By changing the proportions, the CCPS network can be tuned for different applications of next-generation wireless networks. Then, we propose a secure transmission policy selection (STPS) mechanism that maximizes the network utility by using the Markov-decision process (MDP). In STPS, the CCPS network jointly selects the best performing physical layer security policy and the parameters of the selected secure transmission policy to adapt to the changing environmental effects. The proposed STPS is realized by reinforcement learning (RL), considering its real-time decision mechanism where agents can decide automatically the best utility providing policy in an altering environment.
 \blfootnote{This work is supported by TUBITAK under Grant 115E827.}
\end{abstract}

\IEEEpeerreviewmaketitle

\section{Introduction}
\begin{figure}[t!]
	
	\centering
	\includegraphics[width=0.9\linewidth]{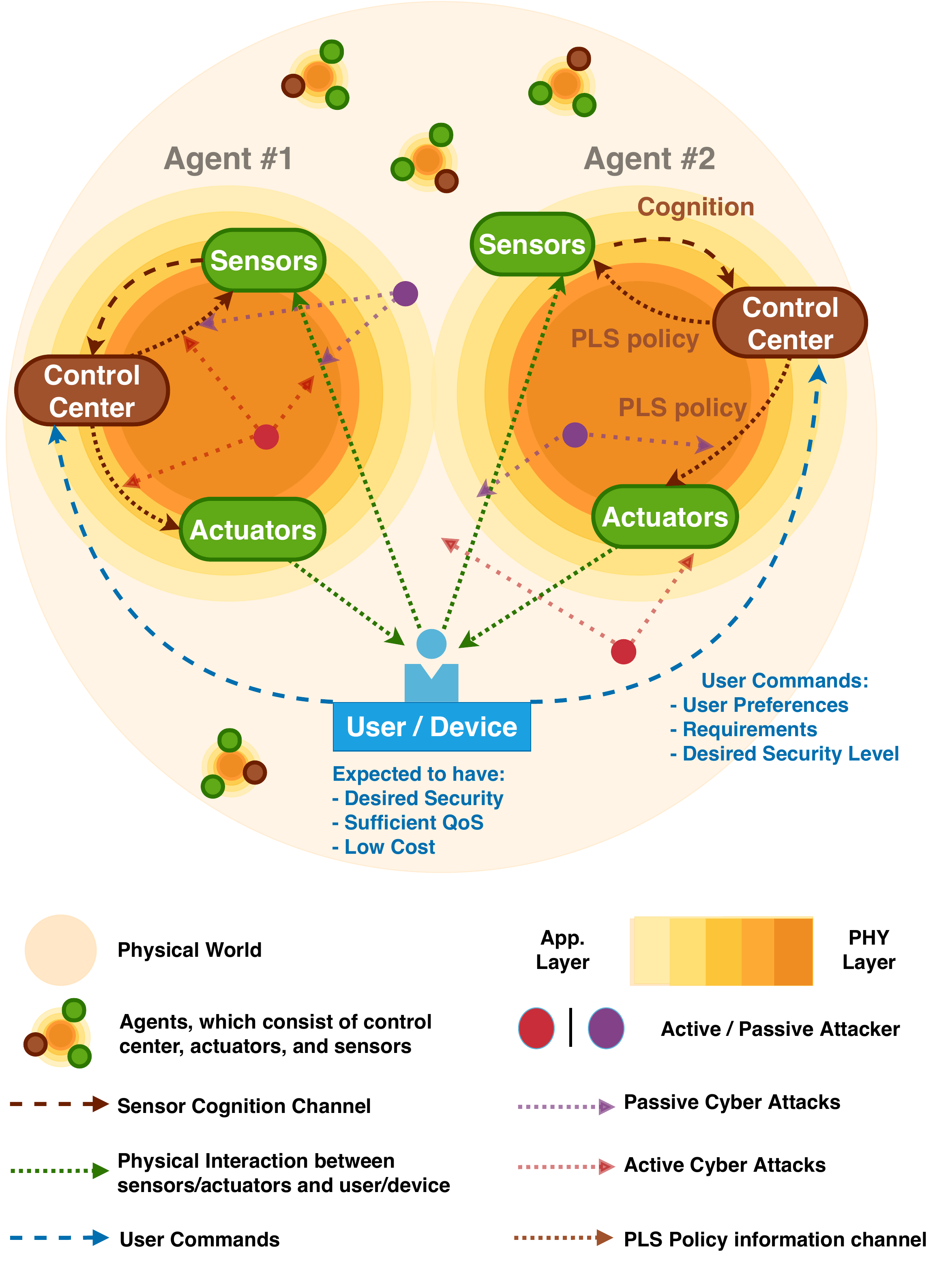}
	\vspace{-0.2cm}
	\caption{System model of a CCPS that consists of multiple agents in layered structure with possible attacks.}
	\label{fig:system model}
	\vspace{-0.3cm}
\end{figure}

Cyber-physical systems (CPS) can a be seen as special form of a distributed system with integrated closed loops for control task among multiple agents, as shown in Fig. \ref{fig:system model}. In this paper, we consider a special type of CPS where all agents communicate their sates and control information over  wireless channels. In this figure, an agent is defined as an individual unit that has a control center, multiple sensors, and actuators. In real-life deployments, a system consists of multiple agents, that interact with each other, such as traffic signs on the road and units inside the vehicles in vehicle-to-everything (V2X), or industrial networks. Typical CCPS have the ability to locally (edge) analyze the environment and make decisions to a limited extent. In this article, we consider agents that utilize reinforcement-learning (RL), where they can analyze the environment, take action in real time. An essential aspect of the environment here is the physical wireless communication channel, which is also essential for monitoring and controlling the security of the communication. On the one hand the communication channel is influenced by the interaction of the agents and on the other hand the physical properties of the communication channel also influence the actions and behavior of the agents.
Hence, the operation of each agent and the coordination between agents depend on the wireless communication infrastructure. However, a wireless network cannot guarantee data security, since radio signals can be intercepted or manipulated by malicious devices \cite{surv_5g}, which may be active or passive attackers, e.g., eavesdroppers, as shown in Fig. \ref{fig:system model}. %\textcolor{black}{Numerous security mechanisms are proposed in the PLS literature, where the body of work has unraveled different security opportunities at physical layer considering various set-ups and attack scenarios\cite{sec_survey}.}

%In many distributed systems, the environment is changing continuously, e.g., additional agents, updated attacker models, increasing noise, and interference. In addition, there are distinct operational requirements of the main applications of CPS, e.g., drone swarms, ultra-reliable low latency communication (URRLC), massive machine type communication (mMTC), smart grid, vehicle-to-everything (V2X) communication, and health networks \cite{demir:2020,topal:2020}. Therefore, CPS also have to adapt to changing environmental conditions to provide continuous security as well as the requirements of the determined application of CPS. In the near future, it is expected that CPS will have cognitive abilities to sense the dynamically changing environments and make decisions based on the updated conditions \cite{topal:2020}. This critical evolution of CPS would be plausible if the agents of CCPS can adapt their operational preferences, e.g., transmission technology, transmit power, security policy, to simultaneously provide high system performance and security \cite{iot}.
In many distributed systems, the environment is also changing continuously, (e.g., additional agents, updated attacker models, increasing noise, and interference). Therefore, CPS also have to adapt to changing environmental conditions to provide continuous security as well as the requirements of the target application. In the near future, it is expected that CPS will have cognitive abilities to sense the dynamically changing environments and make decisions based on the updated conditions \cite{topal:2020}. This critical evolution of CPS would be plausible if the agents of CCPS can adapt their operational preferences, e.g., transmission technology, transmit power, and security policy, to simultaneously provide the desired system performance and security \cite{iot}.

{\color{black}
Numerous security mechanisms are proposed in the physical layer security (PLS) literature, where the body of work has unraveled different security opportunities at the physical layer considering various set-ups and attack scenarios\cite{sec_survey}. The modus operandi in this body of work is maximizing the security performance of a single PLS policy by optimizing the power-frequency-antenna resources under a time invariant channel and attack model. Considering a CCPS framework composed of many agents equipped with various resources, utilizing a single security mechanism with a constant purpose would create a bottleneck for the security and the reliability of the CCPS networks. Instead of focusing on a single specific security scenario, as given in our previous work \cite{topal:2020}, we extend our perspective, and try to answer two large-picture questions for the secure CCPS framework: \textit{"What is the best available security policy and configuration for the CCPS agents under sensed channel conditions?"} and \textit{"Besides security, how can we measure the physical layer performance of a CCPS network?"} 

As an answer to the first question, we propose a decision mechanism, where the best performing PLS policy with best performing transmission configuration is aimed to be selected from the available set of security policies, as shown in Fig. \ref{fig:utility_scheme}. More specifically, we define a control center that is mainly collecting channel data and deciding the best strategy for the agents. The units are designed specifically for sensing, adapting, and configuration located in the control center, where they collect information about ambient risks and environmental changes from sensors, estimating the risks, and communication resources. They also configure the physical layer parameters of the entire system based on agent preferences. Sensors and actuators are then informed by the control center to update transmission parameters. Since the various PLS policies are already available in the literature, the selective nature of the algorithm provides the adaptability for the CCPS network. 

	\begin{figure}[t!]
	\centering
	\includegraphics[width=1\linewidth]{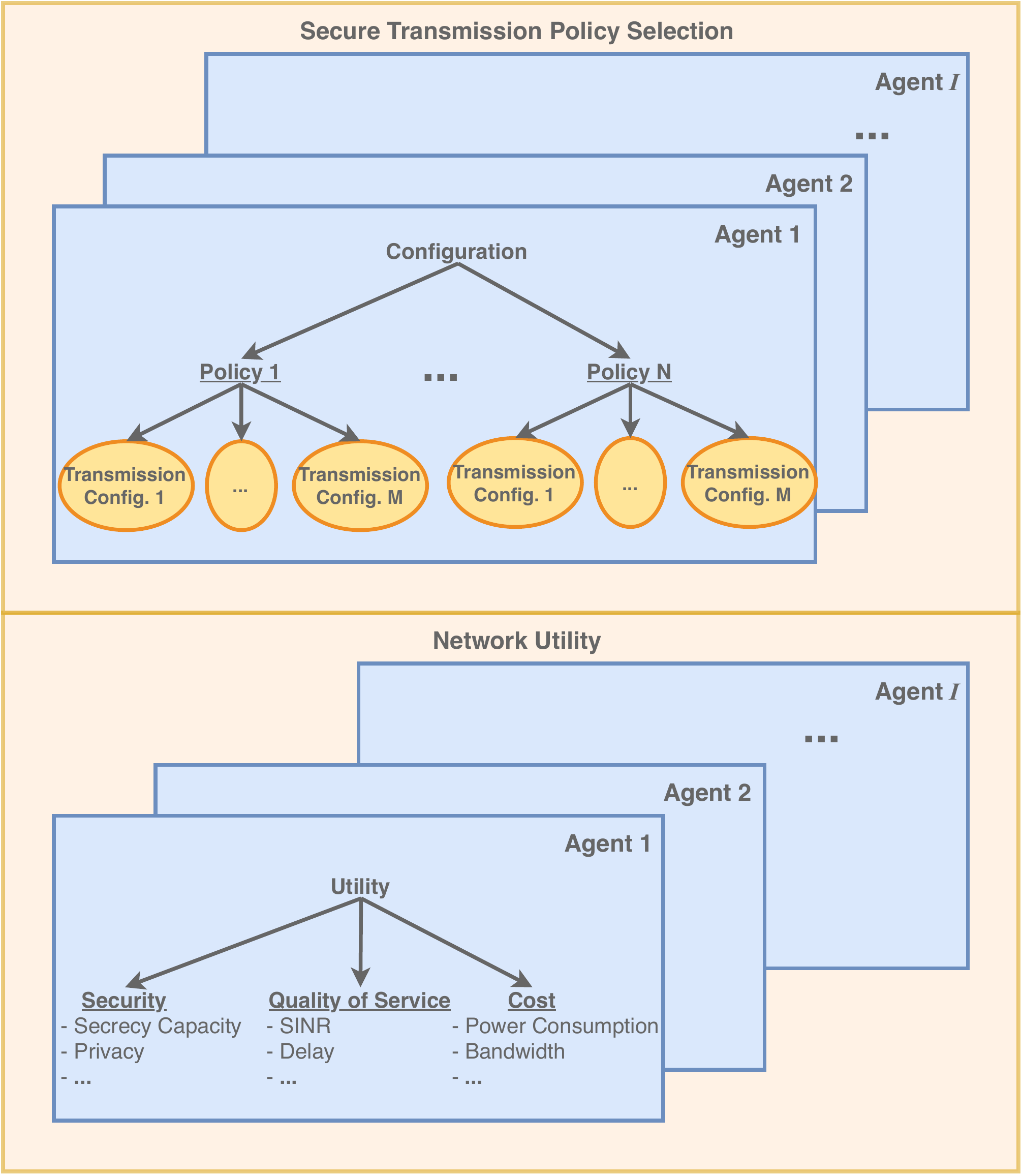}
	\vspace{-0.2cm}
	\caption{Proposed secure transmission policy selection and designed utility metric with its sub-dimensions for each agent.}
	\label{fig:utility_scheme}
	\vspace{-0.3cm}
\end{figure}

As an answer to the second question, we define a user-centric utility as the main performance metric. The utility is the weighted average of the security, quality of service (QoS), and cost dimensions of the CCPS performance. \textcolor{black}{These dimensions include various system performance metrics after classifying into three main dimensions, as shown in Fig. \ref{fig:utility_scheme}}. The weights of each dimension are determined by the requirements imposed by users. \textcolor{black}{Since there are distinct operational requirements of the main applications of CPS, e.g., drone swarms, ultra-reliable low latency communication (URRLC), massive machine type communication (mMTC), smart grid, vehicle-to-everything (V2X) communication, and health networks \cite{demir:2020,topal:2020}, }%Considering the various applications that planned to be provided by the beyond 5G networks, 
we can tune the weights to perform best under the sensed channel conditions and available resources. However, separately selecting the best performing policy for each agent in the network may not provide the best performing network results due to the interference resulted from other agents and the required operational costs. Instead of using individual utilities for each agent in the network, in this work, we propose mixing the utilities of each agent with correct proportions to obtain a single \textcolor{black}{network} utility value that symbolizes the total performance of the CCPS network. To obtain the \textcolor{black}{network} utility, we consider two different operational structures of CCPS networks. In the first structure, the agents calculate their utilities and select the best performing PLS policy individually and locally. In the second structure, agents are connected to a central control unit, where their \textcolor{black}{network} utility is calculated by \textcolor{black}{averaging } %the weighted sum of 
their individual utilities. Our contributions in this work can be listed as below. }

%the operational performance of this utility concept in CCPS is not thoroughly studied yet in the literature, especially for multi-agent CCPS networks. More specifically, the agents may have different priorities and resources, where  an agent selecting individually best-performing policy may deteriorate the utility of other agents considering the interference and resource utilization. In this case, a joint utility structure with an adaptable decision manager would be required to optimize the performance of the framework. 

%In order to fill this gap, the existing CCPS framework, given in \cite{topal:2020}, should be adapted for multi-agent systems for changing environmental conditions. Besides, a real-time decision methodology should be studied for long-term operations. 
\begin{itemize}
    \item We propose an adaptive PLS policy selection and transmission configuration mechanism, \textcolor{black}{which is called secure transmission policy selection (STPS),} for a multi-agent physical layer security framework using an Markov decision process (MDP). The proposed mechanism is inherently adaptable, where it may maximize the individual rewards or a joint reward respectively considering selfish and cooperating agents.
    \item The reward of the individual agents is calculated by a utility function, which is a weighted sum of the QoS, security, and cost dimensions. By multiplying the utility function with time-dependent characteristics, such as degrading battery, we propose a more applicable reward function considering the CCPS frameworks.
    \item We simulate the proposed mechanism with an RL-based setup considering different beyond 5G applications and environment characteristics. Simulation results of the proposed mechanism are compared with the only security-based and only QoS-based policy selection mechanisms to demonstrate the trade-offs between  QoS, security, and cost dimensions. 
\end{itemize}

Overall, we provide an adaptive CCPS framework that is able to make the best \textcolor{black}{STPS} for each agent to maximize its utility, which is defined as the weighted sum of QoS, security, and cost dimensions, based on time-dependent changing environment and application-based requirements.
%\vspace{-0.2cm}
\subsection{Related Works}
The PLS policies and their performance criteria are provided in \cite{sec_survey}. One way to secure the physical layer transmission is a resource allocation at the transmitter node \cite{resource}. Other PLS policies are obtained by the different signal processing methods, such as beamforming and precoding \cite{beamf, beamf2}. Since the message is pre-coded in line with the receiver's channel, the performance of any other node at a different position would decrease \cite{IoTBeam}. Antenna-selection or node-selection based PLS policies are other types of physical layer defense mechanism \cite{hybrid_1}. The joint utilization of different PLS policies is previously considered in \cite{hybrid_1}, and \cite{hybrid_2}. In \cite{hybrid_1}, the authors propose antenna selection with beamforming in MIMO networks. However, these works consider utilizing the same PLS policies in changing conditions. As a difference, we consider that a set of PLS policies are available for the CCPS framework. Similar to our previous work \cite{topal:2020}, this methodology can be categorized under a novel PLS policy selection umbrella in the PLS literature.

Another difference from the aforementioned works is that here, we consider multi-agent security, where the total utility is aimed to be maximized. Similar to our study, in \cite{CPS_game}, the authors propose a game-theoretic physical layer security mechanism, where a model-based predictive control system is utilized to connect the cyber and physical layers of the autonomous systems.  In \cite{NOMA_cps}, the authors consider non-orthogonal-multiple-access (NOMA) to establish a secure wiretap model and optimize the user power allocation coefficients to maximize the secrecy. In \cite{cognitive_1}, the security of CCPS agents is established by the optimal power allocation and power splitting at the secondary transmitter under secrecy constraints. In \cite{cognitive_2},  the medium access probability and transmission power of secondary transmitters are jointly optimized to maximize the security of all network nodes. In \cite{cog_3}, the authors propose a beamforming design for a two-way cognitive radio (CR) Internet of Things (IoT) network aided with the simultaneous wireless information and power transfer (SWIPT). While these works consider the performance of a single method under stable channel characteristics, our work provides an adaptable joint power and PLS policy selection in changing environmental conditions, addressing the corresponding gap in the literature. 

\vspace{-0.2cm}
\subsection{Organization}
In the following section, we detail the calculation of utility dimensions from the physical layer parameters. In Section III, we provide the MDP-based \textcolor{black}{STPS} mechanism and the proposed utility function. In Section IV, we present the simulation parameters and results of the proposed mechanism. In Section V, the paper is concluded.

	\begin{figure*}[ht!]
		\centering
		\includegraphics[width=0.9\linewidth]{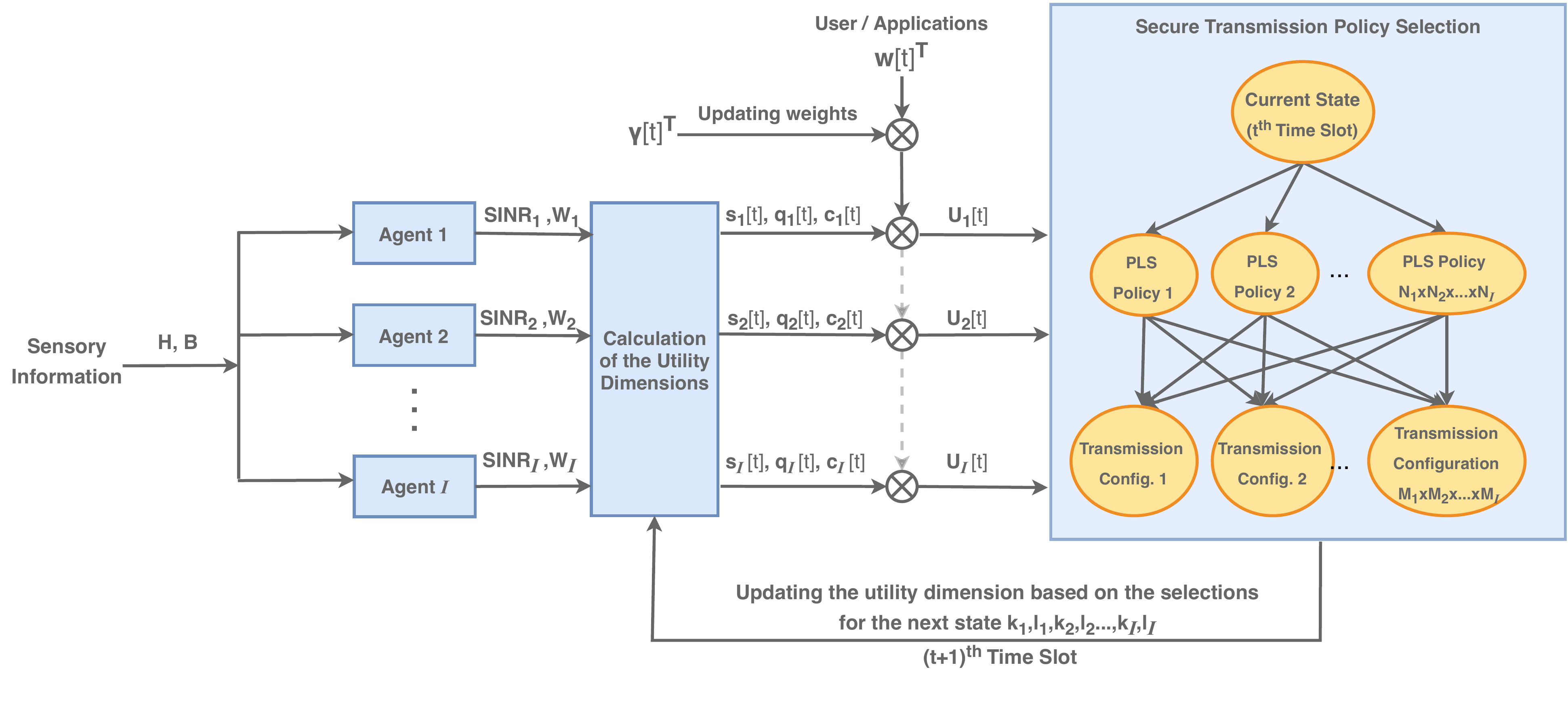}
		\caption{The block diagram of the utility calculation with the proposed MDP-based \textcolor{black}{STPS}.}
		\vspace{-0.3cm}
		\label{fig:RL model}
	\end{figure*}

\color{black}
\vspace{-0.2cm}
\section{Calculating the dimensions of utility}
In our system model, we assume that the CCPS network consists of  $I$ agents, where $i$ denotes the index of an agent. Due to different requirements and environmental conditions, each agent may have different sets of available PLS policies and transmission configurations, which are denoted as $\mathcal{P}_i$ and $\mathcal{R}_i$ for the $i^\text{th}$ agent, respectively. These sets can be stated as $\mathcal{P}_i = \{P_i^{1}, P_i^{2}, \dots P_i^{N_i}\}$ and $\mathcal{R}_i = \{R_i^{1}, R_i^{2}, \dots, R_i^{M_i}\}$, where $N_i$ is the number of available PLS policies and $M_i$ is the number of available transmission configurations for the $i^\text{th}$ agent. When we consider the transmission time slot of a frame, $t$ is the time slot index, where $t = 1, 2, \dots T$. Due to the impact of increasing time slots, we have to distinguish which STPS is decided at time slot $t$ from the sets of $\mathcal{P}_i$ and $\mathcal{R}_i$ for each agent, where $i = 1, 2, \dots, I$. Under these circumstances, $\mathcal{K}$ and $\mathcal{L}$ are the sets of chosen PLS policies and transmission configurations by agents at $t$, where $\mathcal{K} = \{P_1^{k_1}[t], P_2^{k_2}[t], \dots, P_I^{k_I}[t]\}$ and $\mathcal{L} = \{R_1^{l_1}[t], R_2^{l_2}[t], \dots, R_I^{l_I}[t]\}$. In these statements, $k_i$ and $l_i$ are indices of the chosen PLS policies and transmission configurations by each agent and these terms should satisfy following conditions
\begin{equation}
    k_i \leq N_i, \qquad l_i \leq M_i, \qquad \forall i \in \{1, 2, \dots I\}.
\end{equation}

%where  The operations given in this section correspond to the  {\color{black} first half of this block diagram.}   We consider $I$ number of independent agents
In Fig. \ref{fig:RL model}, the block diagram of the proposed system model is given. In this paper, we assume that the agents are independent and each of the agents consists of a transmitter and a receiver unit. We will denote the transmitter and the receiver of the $i^\text{th}$ agent respectively as $a_i$ and $b_i$. 
The number of antennas of the transmitter node of the $i^{th}$ agent is denoted as $W_i$. We assume that each receiver node is equipped with a single antenna. The eavesdropper is also assumed to be equipped with a single antenna. The channel fading coefficient vector between the nodes $c_i$ and $d_j$ is expressed by $\mathbf{h}_{c_id_j}=\left[h_{c_id_j}(1), h_{c_id_j}(2), \cdots, h_{c_id_j}(W_i)\right]^T$, and $h_{c_id_j}(\omega) \sim \mathcal{CN}(0,\frac{1}{D_{c_id_j}})$, where $c\in\{a,b\}$, $d\in\{a,b\}$, $i,j\in\{1,2,\ldots,I\}$ and $i\neq j$. $\mathbf{p}^T$ denotes the transpose of the vector $\mathbf{p}$. We consider three different PLS methodologies: subcarrier-based artificial noise (SC-AN), full-duplex artificial interference (FD-AI), and artificial noise (AN) with supported beamforming. We also consider the case of only beamforming (B), hence the agents may utilize four different PLS policies. The signal models for each of the PLS policies are detailed in the link\footnote{\texttt{The operations to obtain dimensions and utility is detailed in  \url{https://github.com/ozanalpt/PLS-toolbox-for-CCPS/The_Guidebook_of_the_PLS_toolbox_for_CCPS.pdf}}}. For fairness, we consider that the message transmission power, $R^s$, from all agents are equal. The signal to interference and noise ratio (SINR) at node $b_i$ in the $t^{\text{th}}$ time slot is expressed by 
\begin{equation}
\text{SINR}^b_{i}[t]= \frac{|\mathbf{v}_i^H\mathbf{h}_{a_i,b_i}|^2 R^s}{\gamma^{k_{i}}_i(\Upsilon_i[t]+{N_0}_i)},
\end{equation}
%\textcolor{red}{Please check and compare with eq (4):}
%\begin{equation*}
%\text{SINR}^b_{i}[t]= \frac{|\mathbf{h}_{a_i,b_i}\cdot %\mathbf{\rho}_{a_i,b_i}|^2 %R^s}{\gamma^{k_{i}}_i(\Upsilon_i[t]+{N_0}_i)},
%\end{equation*}
where $\Upsilon_i$ denotes the interference from other agents, ${N_0}_i$ denotes the noise variance at the  receiver node of the $i^\text{th}$agent, and $\gamma^{k_{i}}_i$ denotes the interference cancellation error coefficient of the selected PLS policy. The transmit beamforming vector for the $i^{\text{th}}$ agent is denoted by $\mathbf{v}_i=\mathbf{h}_{a_i,b_i}$.  The interference from other agents can be divided into parts as 
\begin{equation}
\Upsilon_i[t]= \sum\limits_{j\neq i} (\Upsilon^s_{ji}[t]+\Upsilon^c_{ji}[t]),
\end{equation}
where $\Upsilon^c_i$ denotes the interference resulted from the message signal transmitted by other agents, and $\Upsilon^a_i$ denotes the interference resulted from the security signal by other agents. The interference from the message signal transmission can be expressed by
\begin{equation}
\Upsilon^s_{ji}[t]= |\mathbf{v}_j^H\mathbf{h}_{a_jb_i}|^2R^s.
\end{equation}
%\textcolor{red}{ What is $\dagger$? Is that the transpose? Is $\mathbf{h}_{a_jb_i} \mathbf{h}_{a_jb_i}^\dagger=||\mathbf{h}_{a_jb_i}||^2$? check eq. (4).} 
The security interference results from the security signal transmitted from other agents, where $\Upsilon^c_{ji}[t]$ denotes the interference from the security signal of $j^\text{th}$ agent. This term differs based on the selected PLS policy. Similarly, considering the SC-AN policy, $\Upsilon^c_{ji}[t]$ becomes
\begin{equation}
\Upsilon^{c,SC-AN}_{ji}[t]= |h_{a_jb_i}|^2R_j^{l_j}[t], 
\end{equation}
where $R_j^{l_j}[t]$ denotes the selected transmission configuration of the $j^\text{th}$ agent. Note that, $\gamma^{k_{i}}_i=2$ for the SC-AN policy. For other policies, $\gamma^{k_{i}}_i=1$. Considering FD-AI policy, $\Upsilon^c_{ji}[t]$  becomes
\begin{equation}
\Upsilon^{c,FD-AI}_{ji}[t]= |h_{b_jb_i}|^2R_j^{l_j}[t]. 
\end{equation}
Considering the AN policy, $\Upsilon^c_{ji}[t]$  becomes
\begin{equation}
\Upsilon^{c,AN}_{ji}[t]= | \alpha_{a_jb_j} \cdot h_{a_jb_i}|^2R_j^{l_j}[t], 
\end{equation}
where $\alpha_{a_jb_j}= \frac{\mathcal{NS}(h_{a_jb_j})}{||h_{a_jb_j}||}$ and $\mathcal{NS}(\zeta)$ denotes the null-space of the vector $\zeta$. Considering the only beamforming policy,
\begin{equation}
\Upsilon^{c,B}_{ji}[t]= 0. 
\end{equation}
Similarly, SINR for the link from the transmitter of the $i^{\text{th}}$ agent to the eavesdropper can be expressed by 
\begin{equation}
\text{SINR}^e_{i}[t]= \frac{|\mathbf{v}_i^H\mathbf{h}_{a_ie} |^2 R_s}{(\Upsilon_{ie}[t]+{N_0}_e)}.
\end{equation}
In this case, $\Upsilon_{ie}[t]$ becomes
\begin{equation}
\Upsilon_{ie}[t]= \sum\limits_{j\neq i} \Upsilon_{je}^s[t]+\sum\limits_{i=1}^{I} \Upsilon_{ie}^c[t], 
\end{equation}
where 
\begin{equation}
\Upsilon_{je}^s[t]=|\mathbf{v}_jh_{a_je}|^2R^s.
\end{equation}
In addition to the interference resulting from the security signal of other agents, the security signal from the considered agent also decreases the performance of the eavesdropper. 

In the following, we detail how physical layer parameters are converted into the dimensions of the individual utility function. 

\subsection{Security}
As a security metric, we utilize secrecy pressure as proposed in \cite{secrecy_pressure}. The main difference between other secrecy metrics is that the location of the eavesdropper is assumed to be unknown. In this way, secrecy pressure shows the security level of the considered medium, instead of focusing on the exact security levels. As a first step, the ergodic secrecy capacity of each point $(x,y)$ of the surface $S$ is obtained. This step provides us with a secrecy map of the surface $S$ for the corresponding security policy. After that, expectation operation is applied over the surface $S$. The location of the eavesdropper is assumed to be a 2D Gaussian random variable as the surface might contain different physical measures in different areas. For example, considering a factory environment, the mean of the Gaussian distribution may be assumed to be the blind spot of the security cameras. The position of the eavesdropper is unknown; therefore, it will be denoted by $(x,y)$. The secrecy capacity for the $i^\text{th}$ agent can be represented by 
\begin{equation}
C^i_{sec}(x,y)=\max\{0,(C^i_B-C^i_E(x,y)),
\end{equation} 
where $C^i_B$ and $C^i_E(x,y)$ are respectively the channel capacities of the link of the $i^\text{th}$ agent's transmitter and receiver, and the link of $i^\text{th}$ agent's transmitter and eavesdropper. The capacity of the $i^\text{th}$ agent's channel can be given as 
\begin{equation}
C^i_B=\frac{1}{2}\log\left(1+{\text{SINR}^b_i}\right), 
\end{equation}
and eavesdropper's channel capacity is 
\begin{equation}
C^i_E(x,y)=\frac{1}{2}\log\left(1+{\text{SINR}^i_e}(x,y)\right). 
\end{equation}
Note that the capacity of a generic $(x,y)$ point is given since the location of the eavesdropper is unknown. Since the secrecy capacity depends on the mutually independent, independent and identically distributed (i.i.d.) channel fading coefficients, the ergodic secrecy capacity can be given as 
\begin{equation}
\tilde{C}^i_{sec}(x,y)= \mathbb{E}\left[C^i_{sec}(x,y)\right],
\end{equation}
where $\mathbb{E}\{\cdot\}$ denotes the expectation operator. Note that the ergodic secrecy capacity is also defined for a generic $(x,y)$ point. Calculating ergodic secrecy capacity for each point on the surface $S$ would result in calculating the secrecy map of the surface.

\begin{figure*}[ht!]
		\centering
		\includegraphics[width=0.9\linewidth]{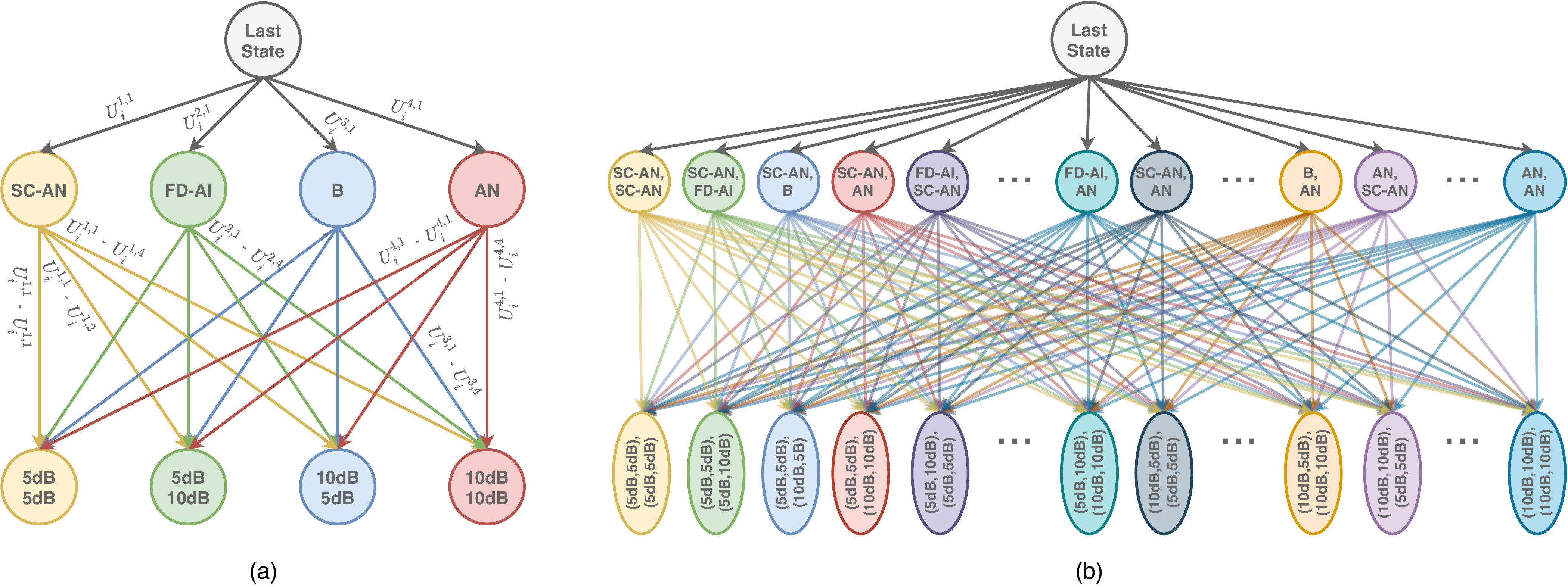}
		\caption{MDP-based PLS policy selection and transmission configuration schemes: (a) individual decision, (b) joint decision}
		\label{fig:RL_single_double}
\end{figure*}

Considering this perspective, we can weigh the ergodic secrecy capacity for each $(x,y)$ generic point by the probability of the eavesdropper's existence at that point. Then, the ergodic secrecy pressure of the surface $S$ of the time slot $t$ can be given by 
\begin{equation}
\Omega_i[t]=\int\int_S \gamma(x,y)\tilde{C}_{sec}(x,y) dx dy, 
\end{equation} 
where $\gamma(x,y)$ is the probability density function (pdf) of the presence of the eavesdropper at a point $(x,y)$ on the surface. In this case, the security of $i^\text{th}$ agent can be obtained by
\begin{equation}
s_i[t]= \frac{\Omega_i[t]}{\max_{j\in\{1,2,\ldots,I\}}[\Omega_j[t]]}.
\end{equation}

\subsection{Quality of Service}
QoS indicates the general performance of the communication link. In the considered setup, we consider the SINR levels at the receiver nodes of each agent as their QoS metric. As described in the security calculation, the SINR levels with respect to each policy and transmission configuration are calculated. Then, they are normalized by the maximum SINR level of the determined STPS. Hence, QoS of the $i^\text{th}$ agent can be described by
\begin{equation}
q_i[t]= \frac{\text{SINR}^b_i[t]}{\max_{j\in\{1,2,\ldots,I\}}[\text{SINR}^b_j[t]]}. 
\end{equation}

\subsection{Cost}
Cost is a measure of the limited resources considering the CCPS framework. The number of active antennas, the selected level of the transmission power, the selected coding scheme are the main elements determining the battery life and the used memory of the CCPS agents. In this work, we consider the cost of the $i^\text{th}$ agent as a weighted sum of the utilized number of antennas and selected transmit power configurations as 

\begin{equation}
c_i[t]=\frac{W_i[t]+R_i^{l_i}[t]}{\max_{j\in\{1,2,\ldots,I\}}[W_j[t]+R_j^{l_j}[t]]}.
\end{equation}
As described in the previous section, the weighted sum of security, QoS, and cost dimensions determine the utility of the agent. The processes explained in this part are conceptualized in  \cite{topal:2020}. Even though this methodology effectively illustrates the general performance of a single-agent, it becomes inefficient for the multi-agent systems, where limited resources, such as frequency bandwidth, battery life, are shared by all agents. Therefore, in the following section, first we obtain a novel performance metric, named as network utility, which models the performance of the CCPS framework considering the joint utilization of the shared resources. Then, we detail the proposed STPS approach for the considered multi-agent framework.  

\section{Adaptive Multi-Layer Security Framework for Multi-Agent Systems}

\subsection{Network Utility Calculation}
During the deployment of an adaptive CCPS framework for multi-agent systems, a proper utility calculation should be done before each transmission, as shown in Fig. \ref{fig:RL model}.  Defining utility is not a straightforward problem; however, it can be calculated as a weighted sum of security, QoS, and cost terms. The calculation of these terms is highly related to the interactions of the agents inside the network, PLS policies, and a proper performance metric. The details about these calculations will be explained in detail in the next section. In short, there are normalized values of security, QoS, and cost, which are denoted as $s_i[t]$, $q_i[t]$, $c_i[t]$.
	
	%, where $\mathcal{K}_{-i}[t]$ and $\mathcal{L}_{-i}[t]$ are the sets of chosen PLS policies and transmit power configurations of all of the agents except the $i^\text{th}$ agent, and can be stated as 
%$\mathcal{K}_{-i}[t] \equiv \mathcal{K} \setminus  P_{i}^{k_i}$ and $\mathcal{L}_{-i}[t] \equiv \mathcal{L} \setminus  P_{i}^{l_i}$.

The importance of these dimensions is fundamentally based on the chosen application of CCPS. However, the roles of the agent may also be distinct from one another; therefore, the weights of the dimensions for each agent may significantly vary. Another factor on the weights is the impact of the active communication time on the weights of each dimension. Due to the rise of power consumption and the reduction of QoS levels of the agents, we try to avoid retransmissions for an increasing number of successful transmission of packets in a frame. To deploy this scheme, we slightly adjust the weights by increasing the weights of QoS and cost dimensions while decreasing the weight of the security dimension for after each successful communication time slot. In these circumstances, we can state a vector of the weights with the impacts of the increasing number of time slots as 
% \begin{equation}
%     \textbf{w}_i[t] =  \begin{bmatrix}
%     \gamma_{s}[t] 
%      \gamma_{q}[t] 
%      \gamma_{c}[t]
%   \end{bmatrix}^T \cdot \mathbb{I}_3 \cdot \begin{bmatrix}
% w_{i,{s}}[t] \\
% w_{i,{q}}[t] \\
% w_{i,{c}}[t]
% \end{bmatrix},
% \label{eq:weights}
% \end{equation}
\begin{equation}
    \textbf{w}_i[t] =  \begin{bmatrix}
    \gamma_{s}[t] 
     \gamma_{q}[t] 
     \gamma_{c}[t]
  \end{bmatrix}^T \circ \begin{bmatrix}
    w_{i,{s}}[t]
    w_{i,{q}}[t] 
    w_{i,{q}}[t]
  \end{bmatrix}^T,
\label{eq:weights}
\end{equation}
where $w_{i,s}[t]$, $w_{i,q}[t]$, and $w_{i,c}[t]$ are the security, QoS, and cost-related weights at $t$ for the $i^\text{th}$ agent, and $\circ$ is the Hadamard product operator. These weights should satisfy  $w_{i,s}[t]+w_{i,q}[t]+w_{i,c}[t]=1$ for $\forall i \leq I$ and $\forall t \leq T$. In \eqref{eq:weights}, $\gamma_{s}[t]$, $\gamma_{q}[t]$, and $\gamma_{c}[t]$ are the impacts of the increasing number of time slots during the transmission of a frame.

 We also assume that there are negative effects on system performance when a different PLS policy is chosen in consecutive time slots. In case of different policies are selected sequentially, the agents may have to turn on/off some of their hardware. This procedure may result in additional delays, and cost; therefore, we also consider these impacts by defining transition values of PLS policies from the policy $k_i$ to $k'_i$, which denotes the next possible values of $k_i$. The transition value is chosen as $\delta_i[t]^{k_i\rightarrow k'_i} = 1$, where $k_i = k'_i$, on the other hand the condition $\delta_i[t]^{k_i\rightarrow k'_i} < 1$ is satisfied, where $k_i \neq k'_i$. A similar formula can be generated for the transmission configuration adjustments of the agents, but we omit these impacts of transmission configuration adjustments to simplify the model. 
%\textcolor{red}{Please give a block diagram how the policies (and which parameters) are modified by the MDP, e.g. give a hint to the examples in the RL Simulations.}
% vector of PLS policies from the policy $k_i$ for the $i^\text{th}$ agent as

% \begin{equation}
%     \textbf{d}_i^{k_i} = [\delta_i^{k_i,1}, \delta_i^{k_i,2}, \dots \delta_i^{k_i,N_i}],
% \end{equation}
% where $\delta_i^{k_i,1}$ is the impact of the changing PLS configuration from $k_i^\text{th}$ policy to $1^{st}$ policy. It should also be noted that $\delta_i^{k_i,k'_i} < 1$ for $k_i \neq k'_i$ and  $\delta_i^{k_i,k_i} = 1$, where $k_i \leq N_i$ $\forall i = 1, 2, \dots I$. 

When we combine dimensions of the utility, weights of each dimension, the impacts of increasing time slots, and transitions to different PLS policies, we can express utility as

\begin{equation}
%\vspace{-0.2cm}
% U_{i}^{k_i,l_i}[t]= \textbf{d}_i^{k_i}(k_i) \times [\textbf{w}_i[t]]^{T} \circ  \begin{bmatrix}
% s_{i}[t]\\
% q_{i}[t] \\
% (1-c_{i}[t])
% \end{bmatrix}^{T},
U_{i}^{k_i,l_i}[t]= \delta_i[t]^{k_i \rightarrow k'_i} \cdot \Bigg( \textbf{w}_i[t]^{T}\times  \begin{bmatrix}
s_{i}[t]\\
q_{i}[t] \\
(1-c_{i}[t])
\end{bmatrix} \Bigg),
\label{eq:utility}
	\end{equation} 
%where $\textbf{d}_i^{k_i}(n)$ is the $n^\text{th}$ element of the $\textbf{d}_i^{k_i}$, where $n \leq N_i$. 

In this utility expression, the cost related term is substracted, since increased cost leads to a smaller utility. The expression \eqref{eq:utility} stands at the center of the CCPS framework, and it will commonly be utilized in the rest of the paper. It should also be remembered that the values of utilities are between $0$ and $1$ due to the normalization of the weights and utility dimensions.

\vspace{-0.2cm}
\subsection{MDP-based Secure Transmission Policy Selection}
%\textcolor{red}{What kind of learning strategy do you apply? Do you use Q-learning. What are the states and what are the actions? I suggest you describe the Q-table of states and actions. What is the reward-function?}

In order to deploy a flexible decision mechanism, an MDP-based \textcolor{black}{Q-learning algorithm} is developed, as shown in Fig. \ref{fig:RL model}. This \textcolor{black}{algorithm} is practiced inside control centers of each agent at each time slot $t$. \textcolor{black}{In the designed Q-learning algorithm, the states are the available PLS policies and transmission configurations, as shown inside the block of STPS in Fig \ref{fig:RL model}.} In the same figure, the current state indicates the last active PLS policy from the time slot $t$. \textcolor{black}{The actions between states are considered as the changing or continuing the current configurations for each transmission timeslot based on the calculated reward values with the aid of utility values.} It should be noted that this model is valid for the individual selections of the agents, and should be updated by combining agent-based procedures for joint policy and transmission configuration selection of agents. 

% \begin{table*}[t!]
% 	\centering
% 	\caption{\textcolor{red}{Q-table}} 
% 	\begin{tabular}{|c|c|c|c|c|c|}
% 		\cline{1-5} 
% 		\textbf{States} & \multicolumn{4}{c|}{\textbf{Actions}}  \\ \cline{2-5}
% 		 (From) & \textcolor{red}{Left} & \textcolor{red}{Mid-Left} & \textcolor{red}{Mid-Right} & 
% 		 \textcolor{red}{Right} \\ \cline{1-5}
% 		Last State  & 0.4 & 0.3 & 0.3 &  \\ \cline{1-5}
% 		SC-AN  & U_{i}^{1,1}[t] - U_{i}^{1,1}[t] & U_{i}^{1,1}[t] - U_{i}^{1,2}[t] & U_{i}^{1,1}[t] - U_{i}^{1,3}[t] &  U_{i}^{1,1}[t] - U_{i}^{1,4}[t]\\ \cline{1-5}
% 		FD-AI &2 & 0.33 & 0.33 & 0.33  \\ \cline{1-5} 
% 		B & 1 & 0.3 & 0.5 & 0.2   \\ \cline{1-5}
% 		AN &2 & 0.4 & 0.4 & 0.2 \\ \cline{1-5} 
% 		(5,5) dB & 0 & 0 & 0 & 0  \\ \cline{1-5}
% 		(5,10) dB & 0 & 0 & 0 & 0 \\ \cline{1-5}
% 		 (10,5) dB & 0 & 0 & 0 & 0   \\ \cline{1-5}
% 		 (10,10) dB & 0 & 0 & 0 & 0  \\ \cline{1-5} 
% 		\end{tabular} %\hspace{0cm}
% 	\label{table:q_table}
% \end{table*}

\textcolor{black}{As we discussed earlier, the number of available PLS policies is $N_i$ and the number of available transmission configurations is $M_i$ for the $i^{th}$ agent. Therefore, the number of states is $N_i + M_i + 1$ after adding the current operational state during the individual operations. It also results in $N_i$ available actions for the first stage of the algorithm and $M_i$ available actions for the second stage for individual decisions. In case of joint decisions by the agents, the number of available actions on the first stage is  $N_1 \times N_2 \times \dots \times N_I$, while the number of available actions for the second stage is  $M_1 \times M_2 \times \dots \times M_I$.} 

With this procedure, an agent decide a STPS based on the calculated utility values by \eqref{eq:utility} for each possible policy and transmission configurations. It also indicates the initial state of the operation at the beginning. In this procedure, there is a reward for each transition, which is calculated with the help of utility values. At the policy selection stage, the rewards are specific values of the calculated utilities with a determined transmission configuration $l_i$ at time slot $t$, i.e., $U_{i}^{1,l_i}[t], U_{i}^{2,l_i}[t], \dots U_{i}^{N_i,l_i}[t]$. The rewards through the second selection stage are calculated using the utility differences of the possible transmission configurations, and the chosen transmission configuration at the policy stage, $l_i$. The rewards from the state of the first PLS policy to the possible transmission configurations of this policy can be written as $U_{i}^{1,1}[t] - U_{i}^{1,l_i}[t] , U_{i}^{1,2}[t]- U_{i}^{1,l_i}[t], \dots U_{i}^{1,M_i}[t]- U_{i}^{1,l_i}[t]$. These values can be positive or negative depending on the values of the utility values and the calculation of its sub-dimensions, which is given in the next section.  

If we update the MDP for a joint selection of PLS policies, the available number of policies and transmission configurations drastically increase. Cooperating agents do not have to operate with the same STPS. In some cases, they should not run with the same STPS, since some of the policies are very disadvantageous and costly if multiple agents operate with them. As a result, we have to extend the MDP procedure scheme for the joint operation. In this case, the available policies at the policy stage will be $N_1 \times N_2 \times \dots \times N_I$. In this setup, each state represents PLS policies of each agent, e.g., a stage indicates two possible PLS policies if we have a joint decision model for two agents. Similarly, the number of available configurations at the transmission configuration selection stage is $M_1 \times M_2 \times \dots \times M_I$, in the case, where $I$ agents jointly make their STPS. In the following section, we utilize RL-based simulations to implement MDP architecture with network utility.
% \color{red}
% In the joint MDP procedure, we need to calculate again utilities of the agents jointly by utilizing individual utility terms given in \eqref{eq:utility}. In this case, we should distinguish the priorities of the agents, e.g., $\textbf{p} = [p_1, p_2, \dots p_I]$ with the condition of $p_1 + p_2 + \dots + p_I = 1$. These values can also be affected for the increased amount of time slots through $T$, but we omit these impacts in the scope of this paper. As a result of the prioritization, the joint utility can be calculated as 
% \begin{equation}
%     U^{\mathcal{K},\mathcal{L}}= \textbf{p} \times \begin{bmatrix} U_{1}^{k_1,l_1}[t], U_{2}^{k_2,l_2}[t],
%     \dots,
%     U_{I}^{k_I,l_I}[t] \end{bmatrix}^{T},
%     \label{eq:weighted_sum_utility}
% \end{equation} 
% where $k_i \leq N_i$, $l_i \leq M_i$, for $\forall i \in \{1, 2, \dots I\}$. Thanks to the prioritization, we are able to assign master and slave agents, and then the MDP procedure can select best joint PLS policies and transmit power configurations with RL after a training.
% \color{black}

\begin{table}[t!]
    \centering
    \caption{\textcolor{black}{RL-based simulation parameters and the decision methodologies for actualizing a STPS}}
    \begin{tabular}{|c|c|}
    	
		\hline
		\textbf{Decision methodology} &  \textbf{Time slot of the agents} \\ \hline
% 		\textbf{Methodology} & \textbf{ID} & \textbf{$1^{st}$ agent} & \textbf{$2^{nd}$ agent} \\\hline
Individual & $t-1$ \\ \hline
Joint & Simultaneously \\
\hline
\multicolumn{2}{c}{}\\
		\multicolumn{2}{c}{\textbf{Simulation Parameters}} \\ \hline
Number of agents, $I$ & 2 \\ \cline{1-2}
     \multirow{2}{*}{Discount rate} & $0.75$ (for individual decisions) \\ & $0.85$ (for joint decisions) \\ \cline{1-2}
     \multirow{2}{*}{Learning rate} & $0.02$ (for individual decisions) \\ & $0.035$ (for joint decisions)  \\ \cline{1-2}
     \multirow{2}{*}{ $\epsilon$} & $0.95$ (for individual decisions) \\ & $0.7$ (for joint decisions)  \\ \cline{1-2}
     \multirow{2}{*}{Number of Training Episodes} & $200$ (for individual decisions) \\ & $300$ (for joint decisions) \\ \cline{1-2}
     Number of time slots, $T$ & 50 \\ \cline{1-2} 
     \multirow{2}{*}{$\delta_i^{k_i\rightarrow k'_i}$} & $0.95$ for $k_i \neq k'_i$\\
     & $1$ for $k_i = k'_i$ \\
     \cline{1-2} 
\end{tabular}
    \label{table:sim_parameters}
\end{table}

\section{Reinforcement Learning Simulations}

In order to analyze the performance of the CCPS security framework for multi-agent systems, an RL-based operation is studied and simulated in MATLAB environment. RL can be easily applied for cyclic system models, including CCPS, thanks to its similar design. 

In the given simulations, we study $I = 2$ agents, which are able to operate individually or jointly. In the scope of this paper, the considered PLS policies can be listed as $\mathcal{P}_1 = \mathcal{P}_2 = \{SC-AN,FD-AI,AN,B\}$ for $N_1 = N_2 = 4$. The determined tuples of transmission configurations for each agent can be written as $\mathcal{R}_1 = \mathcal{R}_2 = \{(5,5),(5,10),(10,5),(10,10)\}$ dB for $M_1 = M_2 = 4$, where $(\cdot,\cdot)$ indicates the transmission power and artificial noise levels of an agent, respectively. A rectangular surface of $4m \times 4m$ is assumed as the CCPS environment. The locations of the transmitter and receiver nodes for the first agent are respectively selected as $(-1,1),(1,1)$, while the transmitter and receiver nodes of the second agent are placed at $(-1,-1),(1,-1)$, respectively. The eavesdropper is located at the center, as $(x_e,y_e)=(0,0)$. The pdf of the eavesdropper's location is 2D Gaussian distribution with variance $1$. The noise variances at the eavesdropper and agents are assumed to be equal to $1$.% The channel coefficients are assumed to be random variables with complex Gaussian distribution as $h_{a_ib_j}\sim \mathcal{CN}(0,\frac{1}{D^2_{a_ib_j}})$.

\begin{figure}[ht!]
\begin{subfigure}{.5\textwidth}
  \centering
  % include first image
  \includegraphics[width=1\linewidth]{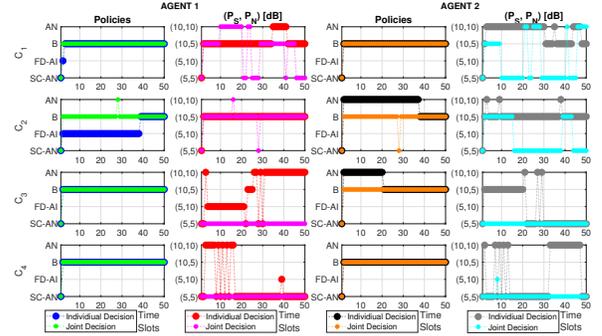}
  \caption{STPS of the agents.}
  \label{fig:agent1_vs_agent2}
\end{subfigure}
\begin{subfigure}{.5\textwidth}
  \centering
  % include second image
  \includegraphics[width=0.9\linewidth]{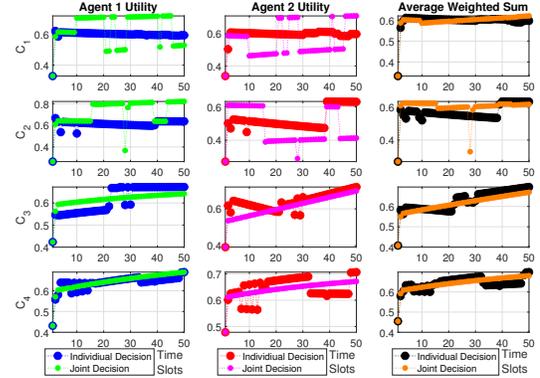}
  \caption{Agent-based utility values and average weighted sum values.}
  \label{fig:utility}
\end{subfigure}
\caption{Simulation results observed with individual and joint decision methodologies for chosen CCPS applications.}
\label{fig:results}
\end{figure}

%\begin{figure*}[ht!]
%\begin{subfigure}{.5\textwidth}
 % \centering
 % % include first image
%  \includegraphics[width=1.17\linewidth,left]{figures/agen1_agent2_policy_power_new.eps}
%  \caption{STPS of the agents.}
%  \label{fig:agent1_vs_agent2}
%\end{subfigure}
%\begin{subfigure}{.5\textwidth}
 % \centering
 % % include second image
%  \includegraphics[width=0.915\linewidth,right]{figures/utility_new.eps}
 % \caption{Agent-based utility values and average weighted sum values.}
 % \label{fig:utility}
%\end{subfigure}
%\caption{Simulation results observed with individual and joint decision %methodologies for chosen CCPS applications.}
%\label{fig:results}
%\end{figure*}

In Table \ref{table:sim_parameters}, we present the RL-based training and simulation parameters, where the future values are taken into account by choosing a discount rate as one while maximizing utility during training RL environment. $T$, which is the number of time slots that affect the behavior of the agents due to changing weights of the utility dimensions, is chosen to be $50$. There are two possible $\epsilon$ values for the designed methodologies of STPS. These values are chosen for maximizing utilities during the operations.    %During our simulations, we assume that we have two agents, e.g. $I=2$, that may cooperate with each other.

{\color{black}
In this paper, we consider four main applications of CCPS, whose weights are given in Table \ref{table:journal_RL_parameters}, and also presented in \cite{topal:2020}. As detailed in the previous section, the utility calculation consists of several parameters, including the vector of weights for each dimension $\textbf{w}_i[t]$, the transition vector $\textbf{d}_i^{k_i}$, impacts of an increasing number of time slots on each dimension denoted as $\gamma_s[t]$, $\gamma_q[t]$, and $\gamma_c[t]$, respectively.  These weights are arbitrarily chosen and can be altered based on usage differences. As the second parameter, each transition from one PLS policy to another creates a utility loss with the $0.05$ ratio; therefore, $\delta_1^{k_i \rightarrow k'_i}  = \delta_2^{k_i \rightarrow k'_i} = 0.95$ when $k_i \neq k'_i$ as given in Table \ref{table:sim_parameters}. This value can also be updated based on various operational environments, which may lead to additional penalties, e.g., delays and distortion for changing PLS policies.}

%\subsection{Simulation Scenarios}
In the simulations, the considered decision methodologies are based on the individual and joint operations of the agents with different Q-learning tables and MDP-based schemes, as shown in Fig. \ref{fig:RL_single_double}. When the agents individually determine their operational PLS policy and transmission configuration, they make this decision based on other agent's STPS, as shown in Table \ref{table:sim_parameters}. On the other hand, the agents simultaneously decide their configurations if they choose joint decision methodology at the same time.  %In this paper, we consider two different approaches for CCPS agents to decide the best PLS policy and transmit power configuration that maximize utility. 

\begin{table}[t!]
	\centering
	\caption{\textcolor{black}{Utility weights for chosen CCPS applications for making a STPS.}} %and the decision methodologies for choosing a PLS policy and transmit power configuration.}
% 	\begin{tabular}{|l|l|}
% 		\hline
% 		T & 50 \\ \hline
% DiscountFactor & 1 \\ \hline
% Epsilon & 0.1 \\ \hline
% EpsilonDecay & 0.01 \\ \hline
% MaxStepsPerEpisode & 50 \\ \hline
% MaxEpisodes & 50 \\ \hline
% StopTrainingCriteria & "AverageReward" \\ \hline
% StopTrainingValue & 13 \\ \hline
% ScoreAveragingWindowLength & 10 \\ \hline
% \multicolumn{2}{c}{} \\
% \multicolumn{2}{c}{(a) RL parameters}
% 	\end{tabular

	\begin{tabular}{|c|c|c|c|c|c|}
		\cline{2-6}
		\multicolumn{1}{c|}{} & \multirow{2}{*}{\textbf{Agent}} & \multicolumn{3}{c|}{\textbf{Utility Weights}} & \multirow{2}{*}{\textbf{Applications}} \\ \cline{3-5}
		\multicolumn{1}{c|}{} &  & $w_{i,s}[1]$       & $w_{i,q}[1]$      & $w_{i,c}[1]$ &  \\ \cline{1-6}
		\multirow{2}{*}{$C_1$} & 1 & 0.4 & 0.3 & 0.3 & \multirow{2}{*}{Drone Swarms} \\ \cline{2-5}
		& 2 & 0.33 & 0.33 & 0.33 & \\ \cline{1-6} 
		\multirow{2}{*}{$C_2$} & 1 & 0.3 & 0.5 & 0.2 & \multirow{2}{*}{URLLC, V2X} \\ \cline{2-5}
		& 2 & 0.4 & 0.4 & 0.2 & \\ \cline{1-6} 
		\multirow{2}{*}{$C_3$} & 1 & 0.2 & 0.3 & 0.5 & \multirow{2}{*}{mMTC, Smart Grid} \\ \cline{2-5}
		& 2 & 0.5 & 0.1 & 0.4 & \\ \cline{1-6} 
		\multirow{2}{*}{$C_4$} & 1 & 0.3 & 0.2 & 0.5 & \multirow{2}{*}{Health Networks} \\ \cline{2-5}
		& 2 & 0.2 & 0.2 & 0.6 & \\ \cline{1-6} 
% 		\multicolumn{1}{|c|}{$C_2$} & 0.3-0.4 & 0.5-0.6 & 0.1-0.2 & URLLC, V2X  \\ \cline{1-5} 
% 		\multicolumn{1}{|c|}{$C_3$} & 0.4-0.5 & 0.1-0.2 & 0.4 & mMTC, Smart Grid \\ \cline{1-5} 
% 		\multicolumn{1}{|c|}{$C_4$} & 0.3 & 0.1-0.15 & 0.55-0.6 & Health networks  \\ \cline{1-5}
% \multicolumn{6}{c}{} \\ 
%  		\multicolumn{6}{c}{(b) Utility weights for chosen CCPS applications} \\ \multicolumn{6}{c}{}
% 		\multicolumn{5}{c}{} \\ \cline{2-5} 
% 		& \multicolumn{2}{c|}{Jamming Power (dB)} & \multicolumn{2}{c|}{Expected Eavesdropper Position}\\ \cline{2-5} q
% 		& $P_{J1}$ & none & $p_{E1}$ & Alice side \\ \cline{2-5}
% 		& $P_{J2}$ & $0$ & $p_{E2}$ & Center\\ \cline{2-5}
% 		& $P_{J3}$ & $5$ & $p_{E3}$ & Bob side\\ \cline{2-5}
% 		& $P_{J4}$ & $10$ & & \\ \cline{2-5}
% 		\multicolumn{5}{c}{} \\ \multicolumn{5}{c}{(b) Attack scenarios for simulations and Eavesdropping} \\ \multicolumn{5}{c}{positions only for SDR-based tests} 
		\end{tabular} %\hspace{0cm}
	
% \\
% \multicolumn{4}{c}{}\\ \hline
%  & \multicolumn{3}{c|}{\textbf{Fayda fonksiyonu ağırlıkları}}\\\hline
% \textbf{Uygulama} & \textbf{Güvenlik} & \textbf{QoS} & \textbf{Maliyet} \\ \hline
% & & &  \\ \hline
% & & &  \\ \hline
% & & & \\ \hline
% & & & \\ \hline
% \multicolumn{4}{c}{}\\
% \multicolumn{4}{c}{(b) Farklı uygulamaların göre}
% \\
% \multicolumn{4}{c}{}\\ \hline
% \multicolumn{2}{|c|}{\textbf{1. birimin önceliği}} & \multicolumn{2}{c|}{\textbf{2. birimin önceliği}}\\ \hline
% \multicolumn{2}{|c|}{0.5} & \multicolumn{2}{c|}{0.5} \\ \hline
% \multicolumn{2}{|c|}{0.1} & \multicolumn{2}{c|}{0.9}  \\ \hline
% \multicolumn{4}{c}{} \\
% \multicolumn{4}{c}{(b) Birimlerin farklı önceliklendirilmesine göre}
	\label{table:journal_RL_parameters}
\end{table}
\begin{table}[t!]
	\centering
	\caption{\textcolor{black}{Q-table that is logged for the weights of the scenario $C_1$ for the second agent when $t=50$. The actions are indicated based on the transitions on the Fig. \ref{fig:RL_single_double}(a), where the condition of $N_i = M_i = 4$ results in $4$ available actions.}} 
	\begin{tabular}{|c|c|c|c|c|c|}
		\cline{2-5} 
		\multicolumn{1}{c|}{} & \multicolumn{4}{c|}{\textbf{Actions}}  \\ \cline{2-5}
		 \multicolumn{1}{c|}{}  & Left & Mid-Left & Mid-Right & 
		 Right \\ \cline{2-5}
		 \multicolumn{1}{c|}{} & \multicolumn{4}{c|}{ \textbf{States} (To)}  \\ \cline{1-5}
		 \textbf{States}  (From) & SC-AN & FD-AI & B & 
		 AN \\ \cline{1-5}
		Last State  & 0.3783 & 0.4145 & 0.5373 & 0.4336  \\ \cline{1-5}
		\multicolumn{5}{c}{} \\ \cline{2-5}
		 \multicolumn{1}{c|}{} & \multicolumn{4}{c|}{ \textbf{States} (To)} \\ \cline{1-5}
		\textbf{States}  (From) & (5,5) dB & (5,10) dB & (10,5) dB & (10,10) dB \\ \cline{1-5}
		
		SC-AN  & 0 & -0.1648 & 0.0011 & -0.2011  \\ \cline{1-5}
		FD-AI & 0  & -0.2121 & 0.0133 & -0.1780  \\ \cline{1-5} 
		B & 0 & 0.0410 & 0.0608 & 0.0523\\ \cline{1-5}
		AN & 0 & -0.1626 & 0.0586 & -0.0997 \\ \cline{1-5} 
% 		(5,5) dB & 0 & 0 & 0 & 0  \\ \cline{1-5}
% 		(5,10) dB & 0 & 0 & 0 & 0 \\ \cline{1-5}
% 		 (10,5) dB & 0 & 0 & 0 & 0   \\ \cline{1-5}
% 		 (10,10) dB & 0 & 0 & 0 & 0  \\ \cline{1-5} 
		\end{tabular} %\hspace{0cm}
	\label{table:q_table}
\end{table}

According to the first decision methodology, each agent makes its decisions that maximize their utility without considering the performance of other agents. The representation of the individual decision methodology given in Fig. \ref{fig:RL_single_double}a for the both of the agents, where $i = 1,2$. In this figure, the rewards of each path are indicated in terms of utility values based on available PLS policies and transmission configurations.  At the policy selection stage,  $(5,5)$ dB is chosen as the reference transmission configuration, where $l_i = 1$. Therefore, the rewards of the transmission configuration selection stage are calculated as the difference between the utility of $(5,5)$ dB transmission configuration and the utility of an available transmission configuration for a selected PLS policy. In this methodology, the rewards, which are calculated with utilities as expressed in \eqref{eq:utility}, are logged in Q-tables, \textcolor{black}{which show the action based rewards,} and utilized in RL-based simulations. \textcolor{black}{An examplary Q-table is given in Table \ref{table:q_table} for the weights of $C_1$ at the end of the assigned frame time, $t= 50$. These values significantly match with  the calculated rewards based on the MDP-based schemes given in Fig. \ref{fig:RL_single_double}}.

In the second scenario, we assume that the agents are able to make joint decisions that maximize the overall system utility. Since the agents do not have to operate at the same STPS, the available number of the PLS policies and the transmission configurations are $N_1 \times N_2$, and $M_1 \times M_2$, respectively. In these circumstances, the MDP-based joint decision methodology is presented in Fig. \ref{fig:RL_single_double}b with $16$ available PLS policies and transmission configurations. %\textcolor{blue}{Since indicating each reward term leads to an unclear presentation, we omit to place most of the reward terms inside the figure.}  

\begin{table*}[ht!]
\scriptsize
      \caption{Average Relative Indicator based on individual decision based policy and transmission configuration selection mechanism with $\epsilon = 0.95$. }
  \begin{tabular}{c|c|c|c|c||c|c|c|c||c|c|c|c||c|c|c|c|}\cline{2-17}
  & \multicolumn{16}{c|}{AGENT 1} \\ \cline{2-17}
       & \multicolumn{4}{c||}{\textbf{Drone Swarms}}  
       & \multicolumn{4}{c||}{\textbf{URLLC}} & \multicolumn{4}{c||}{\textbf{mMTC}}  
       & \multicolumn{4}{c|}{\textbf{Health Networks}} \\ \cline{2-17}
      & \multicolumn{4}{c||}{\textbf{Average relative indicator}} & \multicolumn{4}{c||}{\textbf{Average relative indicator}} & \multicolumn{4}{c||}{\textbf{Average relative indicator}} & \multicolumn{4}{c|}{\textbf{Average relative indicator}} 
            \\ \cline{2-17}

        & Utility & Sec. & QoS & Cost &
       Utility & Sec. & QoS & Cost &
       Utility & Sec. & QoS & Cost &
       Utility & Sec. & QoS & Cost 
       \\ \cline{2-17}
      \cellcolor{yellow!15}\textbf{SCAN}  & \cellcolor{green!100}241\% & \cellcolor{green!55}56\% &  
      \cellcolor{green!100}164\% & 
      \cellcolor{green!50}49\% &  \cellcolor{green!100}188\% & \cellcolor{green!75}73\% & \cellcolor{green!100}161\% & \cellcolor{green!60}61\% & \cellcolor{green!100}418\%  & \cellcolor{green!20}22\% &  \cellcolor{green!100}132\% & \cellcolor{green!35}34\% &  \cellcolor{green!100}478\% & \cellcolor{green!20}22\% & \cellcolor{green!55}54\% & \cellcolor{green!25}23\% 
      \\ \cline{2-17}
       \cellcolor{green!15}\textbf{FD-AI}  & \cellcolor{green!100}150\%  & 
       \cellcolor{green!20}16\% & 
       \cellcolor{green!95}93\% & \cellcolor{green!50}49\% & \cellcolor{green!100}113\% & \cellcolor{green!30}29\% & \cellcolor{green!90}91\% & \cellcolor{green!60}61\% &  \cellcolor{green!100}280\% &  \cellcolor{red!20}8\% & \cellcolor{green!70}69\% & \cellcolor{green!35}34\% & \cellcolor{green!85}325\% & \cellcolor{red!20}22\% & \cellcolor{green!20}12\% & \cellcolor{green!25}23\%  
       \\ \cline{2-17}
          \cellcolor{blue!15}\textbf{B}  & \cellcolor{yellow!50}0\% & \cellcolor{red!20}1\% & \cellcolor{yellow!50}0\% & \cellcolor{green!100}99\% & \cellcolor{red!20}3\% & \cellcolor{green!20}9\% &
          \cellcolor{red!20}1\% &
          \cellcolor{green!100}122\% &  \cellcolor{green!20}7\% & \cellcolor{red!20}22\% & \cellcolor{red!20}12\% & \cellcolor{green!70}69\% & \cellcolor{green!20}13\% & \cellcolor{red!35}35\% & \cellcolor{red!40}42\% & \cellcolor{green!45}46\% 
          \\ \cline{2-17}
         \cellcolor{red!15}\textbf{AN}  & \cellcolor{green!100}105\% & \cellcolor{red!20}4\% & \cellcolor{green!55}56\% & \cellcolor{green!50}49\% & \cellcolor{green!70}71\% & \cellcolor{green!20}6\% & \cellcolor{green!55}54\% & \cellcolor{green!60}61\% & \cellcolor{green!100}210\% & \cellcolor{red!25}24\% & \cellcolor{green!40}37\% & \cellcolor{green!35}34\% & \cellcolor{green!100}246\% & \cellcolor{red!40}37\% & \cellcolor{red!20}10\% & \cellcolor{green!25}23\%   
         \\ \cline{2-17}  \multicolumn{17}{c}{} \\
         \multicolumn{1}{c}{ }& \multicolumn{4}{c}{\textbf{(a)}} &  \multicolumn{4}{c}{\textbf{(b)}} & \multicolumn{4}{c}{\textbf{(c)}} &  \multicolumn{4}{c}{\textbf{(d)}} \\
          \multicolumn{17}{c}{} \\ \cline{2-17}
         & \multicolumn{16}{c|}{AGENT 2} \\ \cline{2-17}
         & \multicolumn{4}{c||}{\textbf{Average relative indicator}} & \multicolumn{4}{c||}{\textbf{Average relative indicator}} & \multicolumn{4}{c||}{\textbf{Average relative indicator}} & \multicolumn{4}{c|}{\textbf{Average relative indicator}} 
            \\ \cline{2-5} \cline{6-9} \cline{10-13} \cline{14-17} 
       & Utility & Sec. & QoS & Cost &
       Utility & Sec. & QoS & Cost &
       Utility & Sec. & QoS & Cost &
       Utility & Sec. & QoS & Cost 
       \\ \cline{2-17}
      \cellcolor{yellow!15}\textbf{SCAN}  & \cellcolor{green!100}142\% & \cellcolor{green!60} 58\% & 
      \cellcolor{green!100} 152\% & 
      \cellcolor{green!50} 49\% &   \cellcolor{green!100}104\% & \cellcolor{green!55}54\% & \cellcolor{green!100}107\% & \cellcolor{green!65}63\% & 
      \cellcolor{green!100} 293\% &  \cellcolor{green!35} 33\% & 
      \cellcolor{green!100} 108\% &  \cellcolor{green!35} 37\% & \cellcolor{green!100} 263\% & \cellcolor{green!20} 18\% & \cellcolor{green!100} 105\% & 
      \cellcolor{green!30} 30\% 
      \\ \cline{2-17}
       \cellcolor{green!15}\textbf{FD-AI}  &  \cellcolor{green!80} 78\% &  \cellcolor{green!20} 16\% & 
       \cellcolor{green!80} 80\% & \cellcolor{green!50} 49\% & \cellcolor{green!50} 49\% & \cellcolor{green!20} 14\% & \cellcolor{green!50} 49\% & \cellcolor{green!60} 62\% &  \cellcolor{green!35} 34\% &  \cellcolor{yellow!50} 0\% & \cellcolor{green!50} 49\% & \cellcolor{green!35} 37\% & \cellcolor{green!100} 167\% & \cellcolor{red!20} 11\% & 
       \cellcolor{green!45} 47\% & \cellcolor{green!30} 30\% 
       \\ \cline{2-17}
          \cellcolor{blue!15}\textbf{B}  & \cellcolor{red!20} 4\% & \cellcolor{red!20} 1\% & \cellcolor{red!20} 5\% & \cellcolor{green!100} 99\% & \cellcolor{red!20} 19\% & \cellcolor{red!20} 3\% & \cellcolor{red!20} 21\% & \cellcolor{green!100} 124\% & \cellcolor{green!20} 8\% & \cellcolor{red!20} 15\% & \cellcolor{red!20} 21\% & \cellcolor{green!75} 74\% & \cellcolor{green!20} 9\% & \cellcolor{red!25} 25\% & \cellcolor{red!20} 22\% & \cellcolor{green!60} 59\% 
          \\ \cline{2-17}
         \cellcolor{red!15}\textbf{AN}  & \cellcolor{green!45} 45\% & \cellcolor{red!20} 4\% & \cellcolor{green!45} 46\% & \cellcolor{green!50} 49\% & \cellcolor{green!20} 22\% & \cellcolor{red!20} 5\% & \cellcolor{green!20} 20\% & \cellcolor{green!60} 62\% & \cellcolor{green!100} 134\% & \cellcolor{red!20} 18\% & \cellcolor{green!20} 20\% & \cellcolor{green!35} 37\% & \cellcolor{green!100} 119\% & \cellcolor{red!25} 27\% & \cellcolor{green!20} 18\% & \cellcolor{green!30} 30\% \\ \cline{2-17} \multicolumn{17}{c}{} \\
         \multicolumn{1}{c}{ }& \multicolumn{4}{c}{\textbf{(e)}} &  \multicolumn{4}{c}{\textbf{(f)}} & \multicolumn{4}{c}{\textbf{(g)}} &  \multicolumn{4}{c}{\textbf{(h)}}
      \end{tabular}
      \label{tab:results_ind}
  \end{table*}
  
   \begin{table*}[ht!]
  \scriptsize
        \caption{Average Relative Indicator based on joint decision based policy and transmission configuration selection mechanism with $\epsilon = 0.7$.}
  \begin{tabular}{c|c|c|c|c||c|c|c|c||c|c|c|c||c|c|c|c|}\cline{2-17}
  & \multicolumn{16}{c|}{AGENT 1} \\ \cline{2-17}
       & \multicolumn{4}{c||}{\textbf{Drone Swarms}}  
       & \multicolumn{4}{c||}{\textbf{URLLC}} & \multicolumn{4}{c||}{\textbf{mMTC}}  
       & \multicolumn{4}{c|}{\textbf{Health Networks}} \\ \cline{2-17}
      & \multicolumn{4}{c||}{\textbf{Average relative indicator}} & \multicolumn{4}{c||}{\textbf{Average relative indicator}} & \multicolumn{4}{c||}{\textbf{Average relative indicator}} & \multicolumn{4}{c|}{\textbf{Average relative indicator}} 
            \\ \cline{2-17}

       & Utility & Sec. & QoS & Cost &
       Utility & Sec. & QoS & Cost &
       Utility & Sec. & QoS & Cost &
       Utility & Sec. & QoS & Cost 
       \\ \cline{2-17}
      \cellcolor{yellow!15}\textbf{SCAN}  & \cellcolor{green!100}266\% & \cellcolor{green!40}39\% &  
      \cellcolor{green!100}180\% & 
      \cellcolor{green!40}41\% &  \cellcolor{green!100}240\% & \cellcolor{green!55}56\% & 
      \cellcolor{green!100}220\% & \cellcolor{green!50}49\% & \cellcolor{green!100}419\%  & \cellcolor{red!20}7\% &  \cellcolor{green!20}19\% & \cellcolor{green!20}16\% &  \cellcolor{green!100}487\% & \cellcolor{red!20}7\% & \cellcolor{green!20}18\% & \cellcolor{green!20}16\% 
      \\ \cline{2-17}
        \cellcolor{green!15}\textbf{FD-AI}  & \cellcolor{green!100}169\%  & 
       \cellcolor{green!20}3\% & 
       \cellcolor{green!100}105\% & \cellcolor{green!40}41\% & \cellcolor{green!100}151\% & \cellcolor{green!20}16\% & 
       \cellcolor{green!100}134\% & \cellcolor{green!50}49\% &  \cellcolor{green!100}277\% &  \cellcolor{red!30}31\% & \cellcolor{red!20}13\% & \cellcolor{green!20}16\% & \cellcolor{green!100}331\% & \cellcolor{red!30}31\% & \cellcolor{red!20}213\% & \cellcolor{green!20}16\%  
       \\ \cline{2-17}
          \cellcolor{blue!15}\textbf{B}  & 
          \cellcolor{green!20}6\% & \cellcolor{red!20}12\% &
          \cellcolor{green!20}6\% &
          \cellcolor{green!80}82\% & \cellcolor{green!20}14\% & \cellcolor{red!20}2\% & \cellcolor{green!20}21\% & \cellcolor{green!100}98\% &  \cellcolor{green!20}7\% & \cellcolor{red!40}41\% & \cellcolor{red!55}55\% & \cellcolor{green!30}32\% & \cellcolor{green!20}14\% & \cellcolor{red!40}41\% & \cellcolor{red!55}55\% & \cellcolor{green!30}32\% 
          \\ \cline{2-17}
         \cellcolor{red!15}\textbf{AN}  & \cellcolor{green!100}119\% & \cellcolor{red!20}14\% & \cellcolor{green!65}66\% & \cellcolor{green!40}41\% & \cellcolor{green!100}105\% & \cellcolor{red!20}4\% & \cellcolor{green!90}89\% & \cellcolor{green!50}49\% & \cellcolor{green!100}207\% & \cellcolor{red!45}43\% & \cellcolor{red!30}29\% & \cellcolor{green!20}16\% & \cellcolor{green!100}252\% & \cellcolor{red!45}43\% & \cellcolor{green!30}29\% & \cellcolor{green!20}16\%   
         \\ \cline{2-17}  \multicolumn{17}{c}{} \\
         \multicolumn{1}{c}{ }& \multicolumn{4}{c}{\textbf{(a)}} &  \multicolumn{4}{c}{\textbf{(b)}} & \multicolumn{4}{c}{\textbf{(c)}} &  \multicolumn{4}{c}{\textbf{(d)}} \\
          \multicolumn{17}{c}{} \\ \cline{2-17}
         & \multicolumn{16}{c|}{AGENT 2} \\ \cline{2-17}
         & \multicolumn{4}{c||}{\textbf{Average relative indicator}} & \multicolumn{4}{c||}{\textbf{Average relative indicator}} & \multicolumn{4}{c||}{\textbf{Average relative indicator}} & \multicolumn{4}{c|}{\textbf{Average relative indicator}} 
            \\ \cline{2-5} \cline{6-9} \cline{10-13} \cline{14-17} 
       & Utility & Sec. & QoS & Cost &
       Utility & Sec. & QoS & Cost &
       Utility & Sec. & QoS & Cost &
       Utility & Sec. & QoS & Cost 
       \\ \cline{2-17}
      \cellcolor{yellow!15}\textbf{SCAN}  & \cellcolor{green!100}226\% & \cellcolor{green!20} 15\% & \cellcolor{green!85} 83\%  & \cellcolor{green!30} 30\% &   \cellcolor{green!100}116\% & \cellcolor{green!20}13\% & \cellcolor{green!55}54\% & \cellcolor{green!30}30\% &
      \cellcolor{green!100}425\% &  \cellcolor{red!20} 8\% & 
      \cellcolor{green!20} 13\% &  \cellcolor{green!20} 16\% & \cellcolor{green!100} 608\% & \cellcolor{red!20} 8\% &
      \cellcolor{green!20} 13\% & 
      \cellcolor{green!20} 16\% 
      \\ \cline{2-17}
       \cellcolor{green!15}\textbf{FD-AI}  &  \cellcolor{green!100} 136\% &  \cellcolor{red!20} 14\% & 
       \cellcolor{green!30} 32\% & 
       \cellcolor{green!30} 30\% & \cellcolor{green!60} 57\% & \cellcolor{red!30} 16\% & 
       \cellcolor{green!20} 10\% & \cellcolor{green!30} 30\% &  \cellcolor{green!100} 283\% &  \cellcolor{red!30} 32\% & 
       \cellcolor{red!20} 20\% & 
       \cellcolor{red!20} 16\% & \cellcolor{green!100} 413\% & \cellcolor{red!20} 8\% & 
       \cellcolor{green!20} 12\% & \cellcolor{green!20} 16\% 
       \\ \cline{2-17}
          \cellcolor{blue!15}\textbf{B}  & \cellcolor{red!20} 9\% & 
          \cellcolor{red!30} 27\% & 
          \cellcolor{red!30} 30\% & \cellcolor{green!60} 60\% & \cellcolor{red!25} 25\% & 
          \cellcolor{red!30} 28\% & 
          \cellcolor{red!40} 42\% & \cellcolor{green!60} 59\% & \cellcolor{green!20} 8\% & 
          \cellcolor{red!40} 42\% & 
          \cellcolor{red!60} 57\% & \cellcolor{green!30} 32\% & \cellcolor{green!20} 15\% & 
          \cellcolor{red!40} 41\% & 
          \cellcolor{red!60} 57\% & \cellcolor{green!30} 32\% 
          \\ \cline{2-17}
         \cellcolor{red!15}\textbf{AN}  & \cellcolor{green!90} 92\% & \cellcolor{red!30} 29\% & 
         \cellcolor{red!20} 7\% & \cellcolor{green!30} 30\% & \cellcolor{green!30} 27\% & \cellcolor{red!30} 30\% & \cellcolor{red!20} 10\% & \cellcolor{green!30} 29\% & \cellcolor{green!100} 213\% & 
         \cellcolor{red!45} 43\% & \cellcolor{red!35} 35\% & \cellcolor{green!20} 16\% & \cellcolor{green!100} 317\% & \cellcolor{red!45} 44\% & \cellcolor{red!35} 35\% & \cellcolor{green!20} 16\% \\ \cline{2-17} \multicolumn{17}{c}{} \\
         \multicolumn{1}{c}{ }& \multicolumn{4}{c}{\textbf{(e)}} &  \multicolumn{4}{c}{\textbf{(f)}} & \multicolumn{4}{c}{\textbf{(g)}} &  \multicolumn{4}{c}{\textbf{(h)}}
      \end{tabular}
      \label{tab:results_joint}
      \vspace{-0.3cm}
  \end{table*}

\section{Results and Discussion}

\subsection{Simulation Results}
\color{black}
In this section, we present the results of simulations considering different CCPS applications determining agent behaviours and capabilities detailed in the previous section.  %for an attacking position of the eavesdropper based on two decision methodologies, chosen utility weights, and the RL training parameters, which are explained in the previous section and presented in Table \ref{table:sim_parameters}.
The results are illustrated in Fig. \ref{fig:results}, Table \ref{tab:results_ind}, and Table \ref{tab:results_joint}, whereas each presentation is applied for the chosen weights given in Table \ref{table:journal_RL_parameters} due to the existence of multiple utility weights with respect to several applications. 

In Fig. \ref{fig:agent1_vs_agent2}, the behavior of two agents is shown for an increasing number of time slots based on two decision methodologies, while the operating STPS are given separately. In these subfigures, readers can follow the changes of each agent, where they make dyna
mic decisions for changing environments. These results show that the agents tend to choose beamforming %and $(10,5)$ dB as transmission configuration 
most of the time to satisfy the operational requirements with individual decision methodology. On the other hand, there is a cooperation between agents with a joint decision methodology, since at least one of the agents chooses $(5,5)$ dB as transmission configuration in most of the cases. %one of the agents operates only with beamforming and the other agent utilizes full-duplex artificial interference with the aid of beamforming. 

The impacts on the performance of these selections in terms of joint utility values are shown in Fig. \ref{fig:utility} for each agent with a comparison of the average weighted sum of agent-based utility values. %This value is calculated by averaging the summation of the individual utility values of both of the agents. 
The results of the weighted sum of utility values show that joint decision methodology provides higher utility values than individual decision methodology most of the time. This observation is expected since the agents operate more adequate PLS policies and transmission configurations with a joint decision methodology rather than individual decisions. %\color{red} The gap between these two methodologies differs from one application to another, and it is separated in mMTC and smart grid-based results, which are cost prioritized applications. 
From our point of view, the amount of fluctuation in individual decision-based results is considerably high compared to joint decision-based outcomes. This situation is observed due to the lack of cooperation between agents. Without cooperation, agents individually try to obtain their maximum utility values, but the selection of one agent may not satisfy the utility maximization conditions for the other agent. This situation is especially valid for health networks and mMTC  applications, as shown in the same figure. On the other hand, if there is a cooperation, the average weighted sum values are more stable.   

%As readers may also realize, the performance of individual and joint decision methodologies rely on chosen CCPS applications. As expected, the utility values resulted from individual decisions are less than the joint decision-based utility values.

%In our point of view, two main observations can appear: The first one is that these agents may choose same PLS policies while operating individually, but they are not tend to choose same policies while making a joint decision. This is a result of a successful cooperation between agents to reduce power consumption while satisfying QoS and security constraints. The second observation is that the agents are tend to decrease their transmit power configuration while making a joint decision comparing to the individual decision methodology. This observation is clearly not true for URLLC applications; however, it is also an expected outcome because of the assigned low weight for the cost dimension for that application.

To observe the secure transmission performance in terms of security, QoS and cost dimensions individually, we focus on Table \ref{tab:results_ind} and Table \ref{tab:results_joint}. In these tables, the average relative indicator results are provided considering the individual and joint decision methodologies for each application and PLS policies. The average relative indicator is identified as the average percentage-based relative differences between the STPS for $T$ and each PLS policies that operate with the transmission configuration $(10,10)$ dB. In other words, these tables show the relative differences for utility and its dimensions when the proposed methodologies are applied instead of operating with a single PLS policy at maximum transmit power levels. Since these tables present dimension-based relative indicators, weights of each dimension given in Table \ref{table:sim_parameters} should also be considered. In Table \ref{tab:results_ind}, readers may observe that the average relative indicator values are consistent with the assigned weights for both of the agents due to the independent and selfish decisions of the agents to satisfy their requirements. On the other hand, the decisions are delivered to maximize the system utility with the joint decision methodology. Therefore, the average relative indicator values of utility in Table \ref{tab:results_joint} are significantly higher than the same group of outcomes in Table \ref{tab:results_ind}. These observations are expected due to the lack of dimension-based constraints while maximizing the utility value. This fact may significantly degrade system performance, and its importance will be discussed in the next section.

\subsection{Discussion and Open Issues}
The lack of dimension-based constraints is the reason behind the inconsistencies between observed high utility values due to the insufficient dimension-based gains. As a result, the maximized utility may not be sufficient to satisfy the specific requirements of the agents. An example is that the most important dimension of the agents is the cost with  weights of $0.5$ and $0.6$, respectively in $C_4$. The results show that the joint decision methodology provides much higher utility than individual decision methodology for this agent; however, individual decision methodology satisfies the cost requirement far better than joint decision methodology. Naturally, this agent should individually operate, while the second agent tries to cooperate with the first agent due to a high amount of utility without any dominant weight for each dimension. Here, a discussion arises on when the average weighted sum of utility values and the average relative indicator should be utilized for deciding on real system deployments. In general, we can state that the agents should jointly decide if their weights are close and there is no dominant weight in Table \ref{table:journal_RL_parameters}, e.g., drone swarms, or there is the dominant QoS weight for both of the agents, e.g., URLLC and V2X applications. On the other hand, the agents should individually operate to maximize their dimension-based requirements, e.g., the agents in mMTC and health network applications. Similar observations can also be made for other applications of CCPS.  

%In order to increase the performance of the joint decision methodology, constraints should be defined during utility maximization in an optimization problem. On the other hand, when the agents individually run, the dimension-based constraints may not be necessary, since the weighted sum of each dimension typically adjusts the utility value regarding the requirements. 
{\color{black}
One predictable open issue is the implementation of the proposed system in a real-time testbed. Since the provided algorithm can be implemented over the software defined radios, considering real-time problems such as hardware impairments and channel estimation errors are needed to be considered for a real-time application. }

Another important open issue is that CCPS may consist of several nodes with various priorities. {\color{black} For example, in a health network, the security and availability of an implant sensor would be more important than the data access point. }  Therefore, there should be a hierarchical order for the agents. We believe that our formulation is sufficiently flexible to define a priority-based definition, which can be extended by adding a multiplication vector to \eqref{eq:utility}.  After this update, we can also model the relationship between the agents based on their importance. {\color{black} For instance, some nodes may sacrifice from their security to enable high QoS of a more important agent.}

\color{black}
\section{Conclusion}
%\vspace{-0.25cm}
In this work, we have proposed a secure transmission policy selection scheme at physical layer based on the Markov decision process for multi-agents in a CCPS environment. The reward of the proposed scheme is termed as the utility, which consists of security, QoS, and cost dimensions. The individual utilities of the agents are consubstantiated to form the joint utility, illustrating the overall performance of the CCPS framework. Reinforcement learning-based simulations are conducted considering two different agent behaviors: the individual policy and transmission configuration selection scheme, and the joint policy and transmission configuration selection scheme. By tuning the weights of the utility dimensions, we analyze the performance of the proposed schemes in different beyond-5G applications. 

The superior utility performance of the proposed policy selection schemes indicates that a single physical layer security method cannot cover all requirements imposed by different applications. As future work, prioritization based hierarchical orders of the agents can be studied. To increase the performance of the adaptivity of the framework, dimension-based constraints should be added to the physical layer security policy and transmission configuration selections.  
\balance
\bibliographystyle{IEEEtran}
\bibliography{main}

\end{document}